# Simulation, Modeling and Prediction of a Pharmacodynamic Animal Tissue Culture Compartment Model by Physical Informed Neural Network

Jiahao Ma (0000-0002-8978-3405, 2190790308@stu.hit.edu.cn)

School of Marine Science and Technology, Harbin Institute of Technology, Weihai, China

**Abstract**:

Compartment models of cell culture are widely used in cytology, pharmacology, toxicology and other fields. Numerical simulation, data modeling and prediction of compartment models can be realized by traditional differential equation modeling methods. At the same time, with the development of software and hardware, Physical Informed Neural Network (PINN) is widely used to solve differential equation models. This work models, simulates and predicts the cell culture compartment model based on the machine learning framework PyTorch with an 16 hidden layers neural network, including 8 linear layers and 8 feedback active layers. The results showed a loss value of 0.0004853 for three-component four-parameter quantitative pharmacodynamic model predictions in this way, which is evaluated by Mean Square Error (MSE). In summary, Physical Informed Neural Network can serve as an effective tool to deal with cell culture compartment models and may perform better in dealing with big datasets.

## Acknowledgments:

This paper is supported by the Animal Tissue Culture Course of Harbin Institute of Technology (Weihai). Thanks Dr. Yan, the teacher of the Animal Tissue Culture Course, for her careful teaching, guidance and help. At the same time, I would like to thank Jingze Liu, a junior student majoring in mathematics at Qingdao University provide some suggestions on the typesetting of mathematical formulas.

# 1. Introduction

Animal tissue culture is widely used in quantitative pharmacology. In the animal tissue culture process, cells often present different states. Among these States, vital state, apoptotic state and necrotic state are common. The mutual transformation and transformation rate of these three states can directly evaluate the cytotoxicity of drugs, which provides a reference for drug research and development and cell biology research (Guchelaar et al., 1998). Many different types of mathematical models have been used to evaluate the state of cells in tissue culture and the toxicological and pharmacological effects of additives, especially the muti-compartment model.

The multi-compartment model is a mathematical model consisting of varying relationships among components and can be represented by differential equations. In 1838, this kind of model was used to distinguish between changes originating from capillary blood and changes originating from extracellular space (Rydhög et al., 1838), which may be the first time for using compartment model in biology research. Subsequently, compartment models were gradually used in cell culture, including pharmacodynamics (Lobo et al., 2002), immunology (Marino et al., 2011) and cell production system (Nakata et al., 2012). In recent years, more and more cell compartment models have been reported, including the two-compartment model for describing plasmid dynamics in the gut (Alderliesten et al., 2022) and the three-compartment model for drug delivery (Becker et al., 2022). The widespread use of compartment models undoubtedly creates a huge demand for tools to deal with them.

Different from the general finite element method (such as the Euler method, Runge-Kutta method, etc.), the Physical Informed Neural Network (PINN) for solving differential equations was not proposed until 1994 (Dissanayake et al.). With the development of Software and Hardware of Scientific computing, the shortcoming that the PINN method needs a lot of calculation is gradually overcome and the advantages are shown by more and more research (Nascimento et al., 2020). With a large number of compartment models being used to model, predict and evaluate the covid-19 pandemic, more and more studies on the application of PINN to compartment model modeling have emerged (Treibert et al., 2022; Long et al., 2021). But only a few research focus on the application of PINN in the field of tissue culture, except a new creative work by Batuwatta-Gamage, which use PINN to predict moisture concentration and shrinkage of a plant cell during drying (Batuwatta-Gamage et al., 2022). In summary, PINN gives researchers a method to deal with modeling prediction and numerical simulation of compartment models in animal tissue culture.

In this work, we model, simulate and predict a widely cited cell culture compartment model introduced by Guchelaar in 1998. The Implementation of PINN is based on the machine learning framework PyTorch with an 8 layers neural network, including four linear layers and four feedback active layers. After the introduction, the data and method part briefly describes our data and methods, especially the mathematical derivation and neural network model. The results section will focus on the process, results and evaluation of our modeling and prediction. After that, the discussion part mainly focuses on the comparison between our work and other work, as well as suggestions for follow-up research. To summarize the whole work, the conclusion gives the overall contribution of this study and the views drawn from it.

## 2. Materials and Methods

2.1 Data Source

Data collection from the graph & table in Guchelaar's research, which is the data from HSB2 cells determined upon continuous incubation with $10^{-6}$ $M$ ARA-C, including the percent number of cell's different states in over 70h.

2.2 Cell state compartment model of animal tissue culture

The compartment model we used is shown in Figure 1, which is a three-component four-parameter classical model. In the model, cells are allowed to transfer from a vital state to an apoptotic and necrotic state, in which the necrotic state is irreversible, and the transferred state can return to vital or further necrosis.

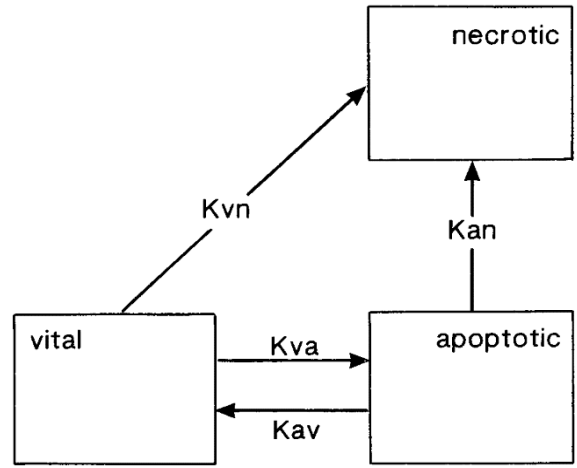

Figure 1 Structure diagram of cell state compartment model
(Guchelaar, 1998)

The differential equation of this model can be written as the differential form of three components with respect to time:

$$\frac{dV}{dt} = -(Kva + Kvn) \cdot V + Kav \cdot A$$

$$\frac{dA}{dt} = Kva \cdot V - (Kav + Kan) \cdot A$$

$$\frac{dN}{dt} = Kvn \cdot V + Kan \cdot A$$

2.3 Construction and Training of Neural Network

The main network structure of this work uses the form of four linear layers and four activation layers, which is shown in Figure 2. Mean Square Error (MSE) is set as the loss function to optimize the training process. The selection of these structures is mainly based on the author's experience in MIT AI: machine learning in healthcare - Johnson program.

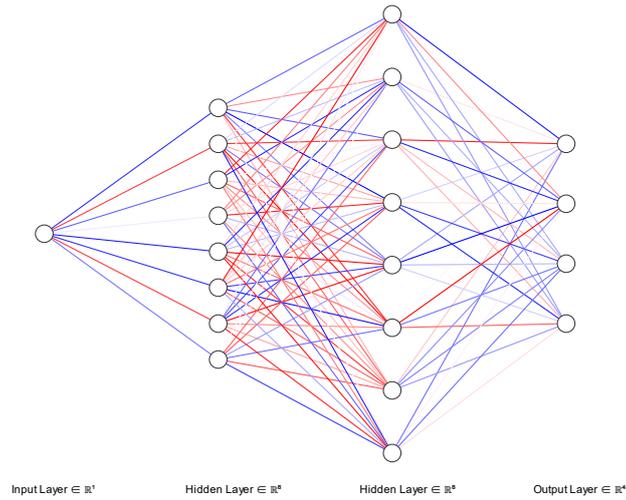

Figure 2 Network Structure Diagram

## 3. Result

### 3.1 Data Source

This part will present the fitting curve and parameters of the training results. After $5*10^4$ rounds of training in 3 minutes, the fitting curve of the demonstration data is shown in the figure below.

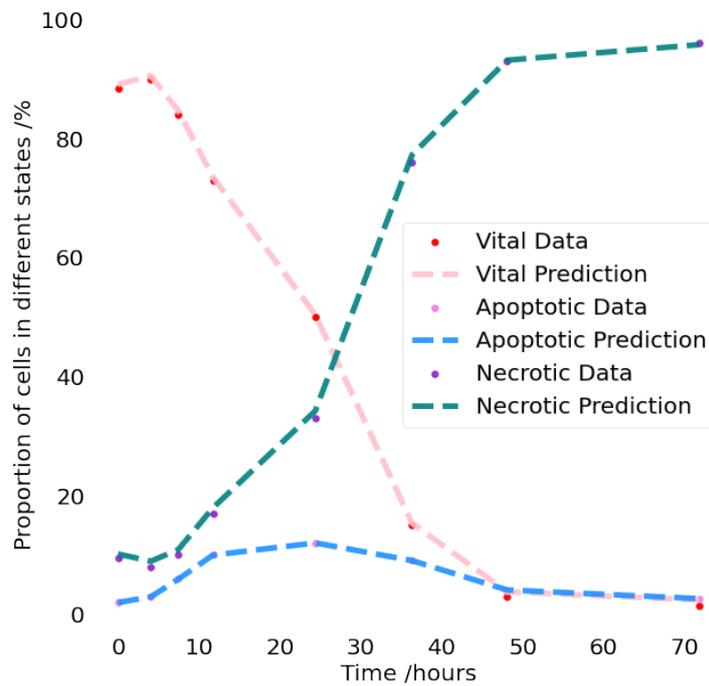

Figure 3 Demonstrate Model Fit Curves

Under the action of the Adam optimizer and Cycliclr learning rate oscillator, the Loss values based on MSE change with the number of training rounds, as shown in Figure 4. The final MSE value is 0.0004853.

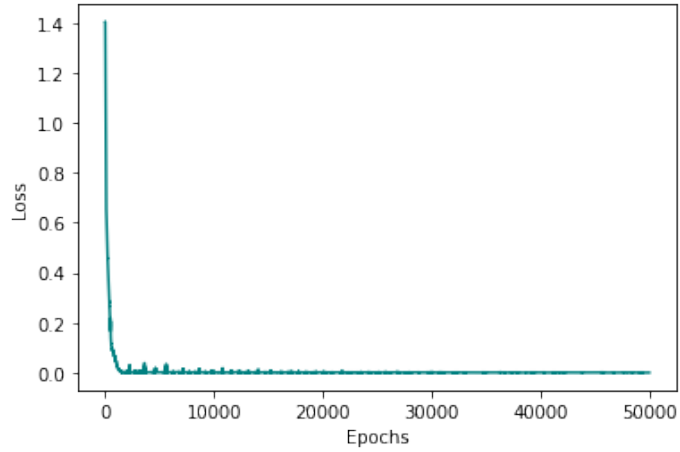

Figure 4 Curve of Loss Values with the Number of Training Rounds

The final model parameters are shown in Table 1.

Table1: The Final Model Parameters

| Name | Kav | Kva | Kan | Kvn |
| --- | --- | --- | --- | --- |
| Value | -0.2233 | 0.0014 | 0.2388 | -0.0081 |

The following Figure 5 uses the parameters obtained from training to conduct numerical simulation on the percentage of each state of cells at different times.

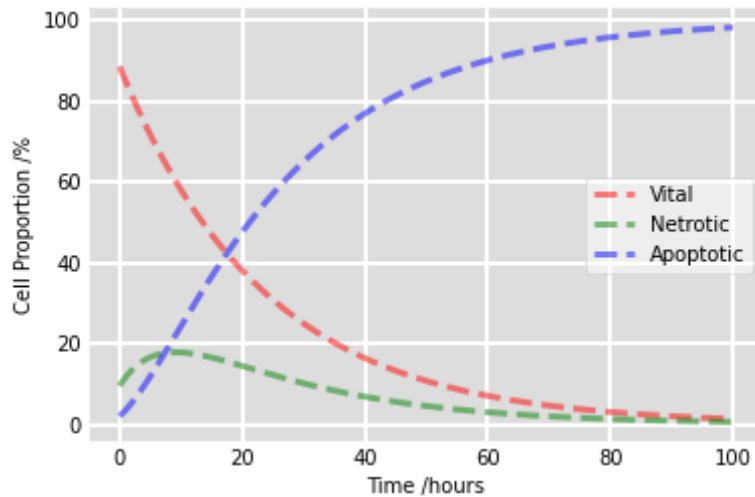

Figure 5 Parameter-driven Numerical Simulation

## 4. Discussion

In recent years, numerous fields have been subverted by artificial neural network technology, including artificial intelligence and machine learning. But because machine learning is a research method rising in recent years, it is not yet fully mature. The use of machine learning methods for cell state modeling is indeed relatively rare. Based on this study, we can see the great potential of Physical Informed Neural Network in animal tissue culture, various cell cultures and pharmacological research.

Compared with related studies, in terms of accuracy, the neural network method of this study is based on hundreds of thousands of rounds of training, and its accuracy is two orders of magnitude higher than that of the original study (Guchelaar et al., 1998). However, from the data set of Batuwatta-Gamage's studies (2022), the small data set limits the play of this method. The method of informed neural networks should be more suitable for larger data sets. Compared with the research on Deep Neural Network (Bianco et al., 2022), PINN is a relatively simple network structure, which makes it more suitable for the cross-field of knowledge-driven modeling and data-driven modeling. In contrast, Deep Neural Networks generally performs well only in data-driven modeling, which limits the application of this method in small data sets.

In some perspective, this is an immature attempt. For further study, larger data sets, more optimal network structure and training methods, and linkage with other machine learning methods are necessary. Specifically, a large number of pharmacodynamic quantitative models and their contrasts can to some extent be evaluated within the same training process, which needs a large dataset with or without lebels. Moreover, Shallow networks and different learning rate oscillators may better cope with overfitting of different data sets as well as training efficiency problems.

## 5. Conclusion

In summary, PINN is a method based on machine learning and artificial neural network. This study shows that it can be used to model, predict and simulate the compartmental models in animal tissue culture, which have a better precision, faster speed, and some scalability.

## Reference

[1] Guchelaar, H. J., Vermes, I., Koopmans, R. P., Reutelingsperger, C. P., & Haanen, C. (1998). Apoptosis-and necrosis-inducing potential of cladribine, cytarabine, cisplatin, and 5-fluorouracil in vitro: a quantitative pharmacodynamic model. *Cancer chemotherapy and pharmacology*, 42(1), 77-83.
[2] Rydhög, A. S., Ahlgren, A., Szczepankiewicz, F., Wirestam, R., Westin, C. F., Knutsson, L., & Pasternak, O. (1838). Joint estimation of free water and perfusion fraction in human brain. *In Proc. Intl. Soc. Mag. Reson. Med* (Vol. 25, p. 2017).
[3] Lobo, E. D., & Balthasar, J. P. (2002). Pharmacodynamic modeling of chemotherapeutic effects: application of a transit compartment model to characterize methotrexate effects in vitro. *AAPs PharmSci*, 4(4), 212-222. https://doi.org/10.1007/s10928-021-09749-w
[4] Marino, S., El-Kebir, M., & Kirschner, D. (2011). A hybrid multi-compartment model of granuloma formation and T cell priming in tuberculosis. *Journal of theoretical biology*, 280(1), 50-62. https://doi.org/10.1016/j.jtbi.2011.03.022
[5] Nakata, Y., Getto, P., Marciniak-Czochra, A., & Alarcón, T. (2012). Stability analysis of multi-compartment models for cell production systems. *Journal of biological dynamics*, 6(sup1), 2-18. https://doi.org/10.1080/17513758.2011.558214
[6] Alderliesten, J. B., Zwart, M. P., de Visser, J. A. G., Stegeman, A., & Fischer, E. A. (2022). Second compartment widens plasmid invasion conditions: Two-compartment pair-formation model of conjugation in the gut. *Journal of theoretical biology*, 533, 110937. https://doi.org/10.1016/j.jtbi.2021.110937
[7] Becker, S., Kuznetsov, A. V., Zhao, D., de Monte, F., & Pontrelli, G. (2022). Model of drug delivery

to populations composed of two cell types. *Journal of theoretical biology*, 534, 110947. https://doi.org/10.1016/j.jtbi.2021.110947

[8] Dissanayake, M. W. M. G., & Phan-Thien, N. (1994). Neural-network-based approximations for solving partial differential equations. *communications in Numerical Methods in Engineering*, 10(3), 195-201. https://doi.org/10.1002/cnm.1640100303

[9] Nascimento, R. G., Fricke, K., & Viana, F. A. (2020). A tutorial on solving ordinary differential equations using Python and hybrid physics-informed neural network. *Engineering Applications of Artificial Intelligence*, 96, 103996.

[10] Treibert, S., & Ehrhardt, M. (2022). A Physics-Informed Neural Network to Model COVID-19 Infection and Hospitalization Scenarios.

[11] Long, J., Khaliq, A. Q. M., & Furati, K. M. (2021). Identification and prediction of time-varying parameters of COVID-19 model: a data-driven deep learning approach. *International Journal of Computer Mathematics, 98*(8), 1617-1632.

[12] Batuwatta-Gamage, C. P., Rathnayaka, C. M., Karunasena, H. C. P., Wijerathne, W. D. C. C., Jeong, H., Welsh, Z. G., ... & Gu, Y. T. (2022). A physics-informed neural network-based surrogate framework to predict moisture concentration and shrinkage of a plant cell during drying. *Journal of Food Engineering*, 111137. https://doi.org/10.1016/j.jfoodeng.2022.111137

[13] Marmolin, H. (1986). Subjective MSE measures. IEEE transactions on systems, man, and cybernetics, 16(3), 486-489.

[14] Bianco, V., Priscoli, M. D., Pirone, D., Zanfardino, G., Memmolo, P., Bardozzo, F., ... & Tagliaferri, R. (2022). Deep learning-based, misalignment resilient, real-time Fourier Ptychographic Microscopy reconstruction of biological tissue slides. *IEEE Journal of Selected Topics in Quantum Electronics*, 28(4), 1-10.